\documentclass[11pt,twoside]{article}

\usepackage{asp2006}
\usepackage{epsf}
\usepackage{psfig}
\usepackage{lscape}

\begin{document}
\title{Identification and Morphologies of Galaxies with Old Stellar Populations at \boldmath $z\sim2.5$}   
\author{Alan Stockton and Elizabeth McGrath}   
\affil{Institute for Astronomy, University of Hawaii, 2680 Woodlawn Drive, Honolulu, HI  96822}    

\begin{abstract} 
We describe a study of morphologies of galaxies with old stellar populations in radio-source fields
at $z\sim2.5$. A significant fraction of these are dominated by {\it disks} of old stars, and none 
we have found so far
has the properties of present-epoch ellipticals.  Recent Spitzer IRAC data confirms that at least one
of our prime examples is definitely not a reddened star-forming galaxy.
\end{abstract}

\section{Introduction}
Detailed studies of galaxies in the local universe and
recent investigations of red-sequence galaxies at high redshifts both indicate that a significant
fraction of massive galaxies formed at very early epochs and within a very short time span.
It has also become increasingly clear that galaxy formation involves feedback effects that are extremely 
difficult to model realistically. As \citet{sil00} has emphasized, ``{\it Ab initio} calculations, 
beginning with inflationary fluctuations that evolve into mature 
galaxies, are currently a theorist's dream and are likely to remain so.''
In other words, models of galaxy formation need strong observational
guidance to constrain the various possible theoretical scenarios.

We have had a program underway for some time now to isolate and investigate, at 
a few specific redshift ranges, galaxies that comprise 
the oldest stellar populations at each 
redshift. Even tiny amounts of residual star formation in a dominant old stellar population
can drastically change the colors of a galaxy at short rest-frame wavelengths. Galaxies that
show no evidence for such star formation
are therefore quite rare, but they offer us the prospect of a clear view of processes that are
likely to be important in massive galaxies generally. Our program is ultimately 
aimed at determining morphologies for a significant sample of such
galaxies at high redshifts.
Morphologies are important because they can preserve direct information 
relevant to formation mechanisms.

We shall use the term ``old galaxies'' (OGs) for galaxies that comprise stellar 
populations that, at their observed epoch, are already old and have virtually no admixture 
of recent star formation. 
We concentrate on identifying OGs that are in groups and clusters
associated with radio sources, for two main reasons: (1.)
radio source fields are likely to include regions of higher-than-average
density in the early universe, in which the very first
massive galaxies are likely to have formed (see, e.g., Best 2000; Barr et al.~2003); and 
(2.) looking for
companions to radio sources {\it at a specific redshift} allows us to choose redshifts for
which the photometric diagnostics from standard broadband filters give the cleanest 
possible separation between old galaxy populations and highly reddened star-forming
galaxies or other possible contaminants. We concentrate here on $z\sim2.5$, for which 
the 4000 \AA\ break falls between the $J$ and $H$ bands.

\section{High-Resolution Imaging}
The first example in this redshift range that we found was 4C\,23.56 ER1 \citep*{sto04,kno97}. 
This galaxy,
with passive evolution alone, would end up as a $\sim2L^*$ galaxy at the present epoch.
Our Subaru adaptive optics (AO) imaging and subsequent 
modelling with {\sc galfit} \citep{pen02} showed it to have very close to an exponential 
profile and a small axial ratio.  The galaxy has every appearance of being an {\it almost
pure disk of old stars.}  Recent {\it HST} NICMOS images confirm these results.

\begin{figure}[!ht]
\plotone{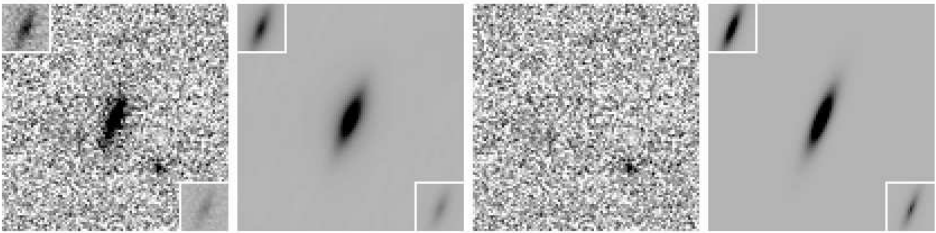}
\caption{Image and models of 4C\,29.28 ER2. The left panel shows the Keck II LGSAO 
$K'$-band image, the 2nd panel shows the {\sc galfit} model, the 3rd panel shows the 
residual of the subtraction of the model from the AO image, and the right panel
shows the {\sc galfit} model without convolution with the PSF. Insets show lower
contrast versions of the images.}\label{4c29img}
\end{figure}

We currently have reasonably good high-resolution (NICMOS or AO) imaging of 6
OGs at $z\sim2.5$, almost all of which are similar in luminosity to 4C\,23.56 ER1.
None of these have the characteristics of normal ellipticals at the present epoch.
We have space here to show only one new example, 4C\,29.28 ER2, which is the
closest to edge-on of several galaxies having low S\'{e}rsic indices (Fig.~\ref{4c29img}).

\begin{figure}[!ht]
\plottwo{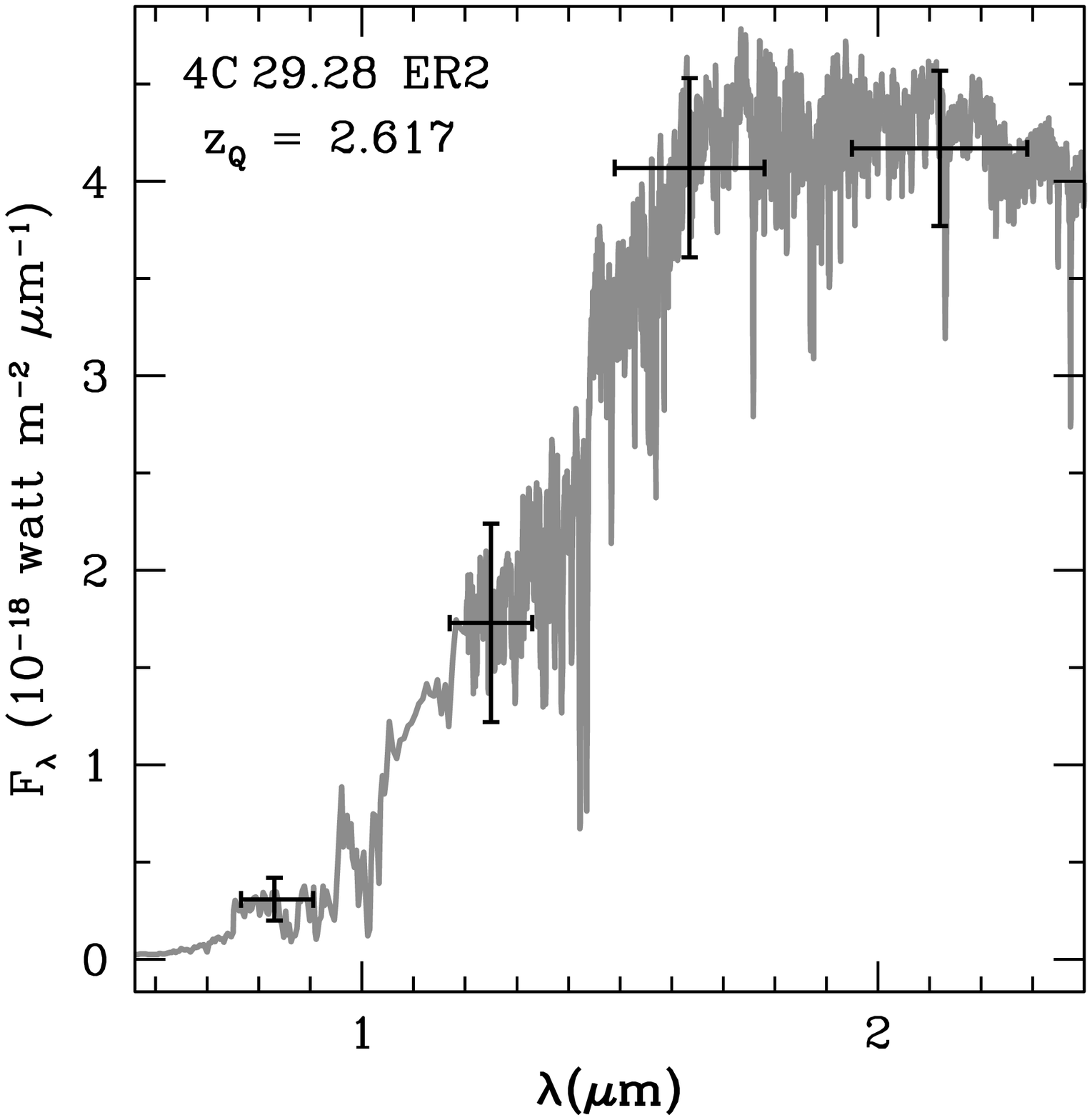}{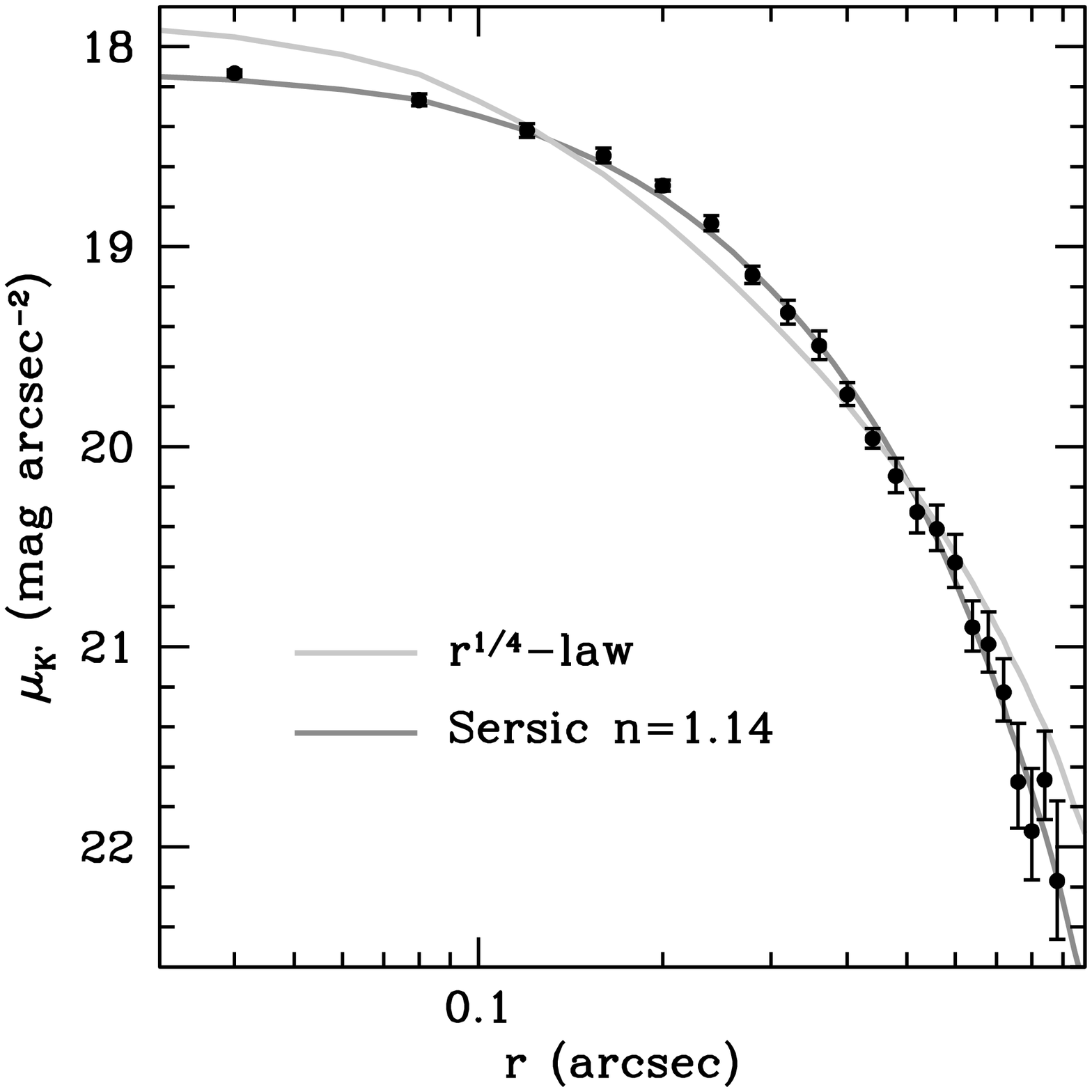}
\caption{SED and radial-surface-brightness profile of 4C\,29.28 ER2. ({\it left panel})---The
observed-frame SED (points with error bars). The gray trace is a 2.1-Gyr-old instantaneous burst
\citet{bru03} model. ({\it right panel})---The radial-surface-brightness profile 
($r$ is the semi-major axis) of the
Keck II LGSAO image shown in Fig.~\ref{4c29img}, with traces showing the best-fit
$r^{1/4}$-law (light gray) and S\'{e}rsic (darker gray) profiles.  
An exponential profile would be virtually identical to
the S\'{e}rsic profile.}\label{4c29}
\end{figure}

The left panel of Fig.~\ref{4c29} shows the spectral-energy distribution (SED) for
4C\,29.28 ER2, which closely matches that for an instantaneous burst model for which
all of the stars were formed when the universe was just $5\times10^8$ years old. The
right panel shows the radial-surface-brightness profile, together with the best-fit
$r^{1/4}$-law and S\'{e}rsic profiles. The profile confirms the visual impression of a
nearly edge-on disk, with no hint of a significant bulge.
\section{Dusty Star-Forming Galaxies?}
At this point, it is natural to ask, ``Couldn't these be, after all, simply dusty star-forming
galaxies?''  Our photometric criteria are designed to minimize this possibility by insisting
on a sharp break in the spectrum, as one sees in the left panel of Fig.~\ref{4c29}.  Nevertheless,
the $H\!-\!K'$ baseline is fairly short, and it could be argued that, within perhaps $2 \sigma$,
one might be able to fit a steeper curve through this region.  In fact, \citet{pie05} apparently simply 
assume that 4C\,23.56 ER1 \citep{sto04} {\it must} be a dusty, star-forming galaxy.

\begin{figure}[!ht]
\plottwo{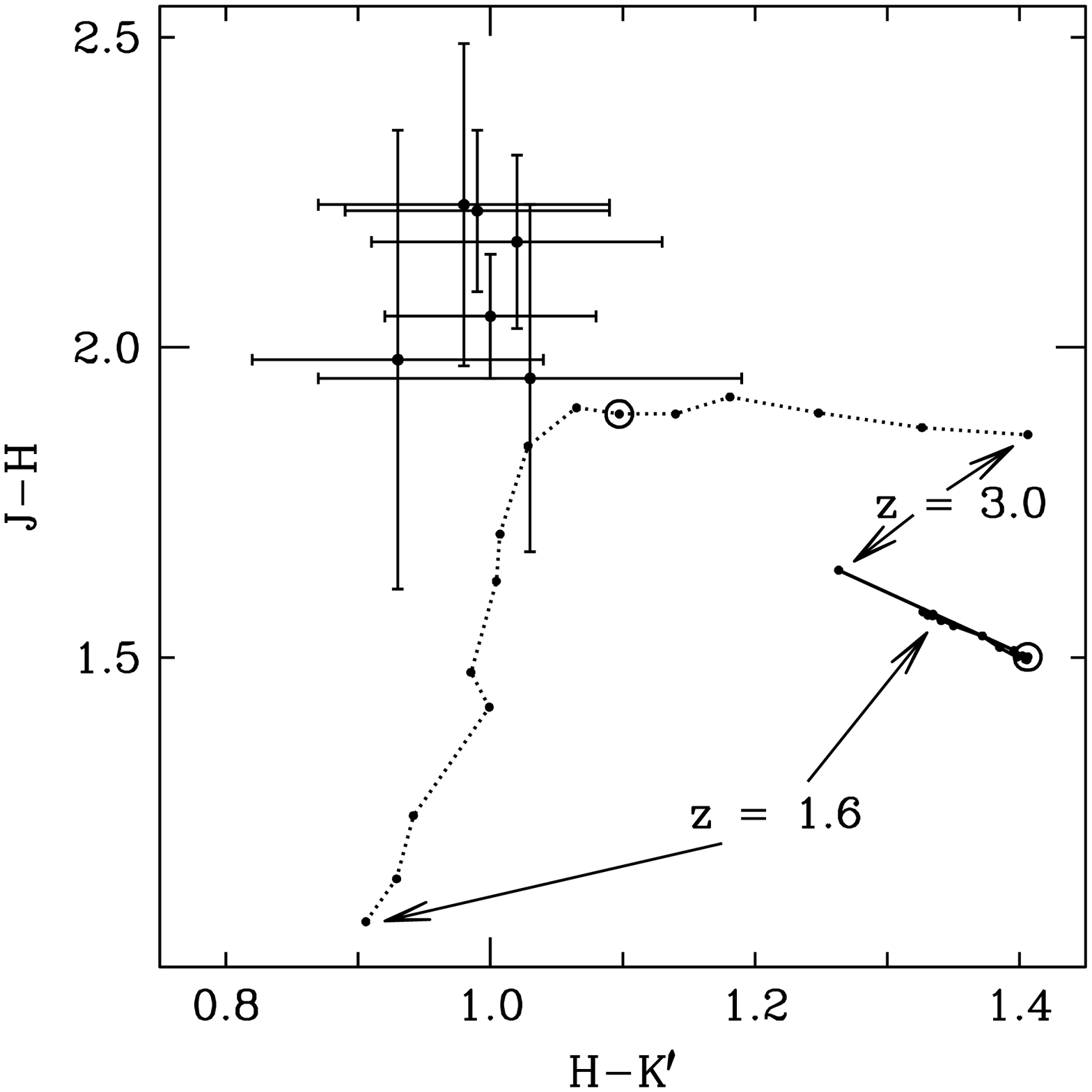}{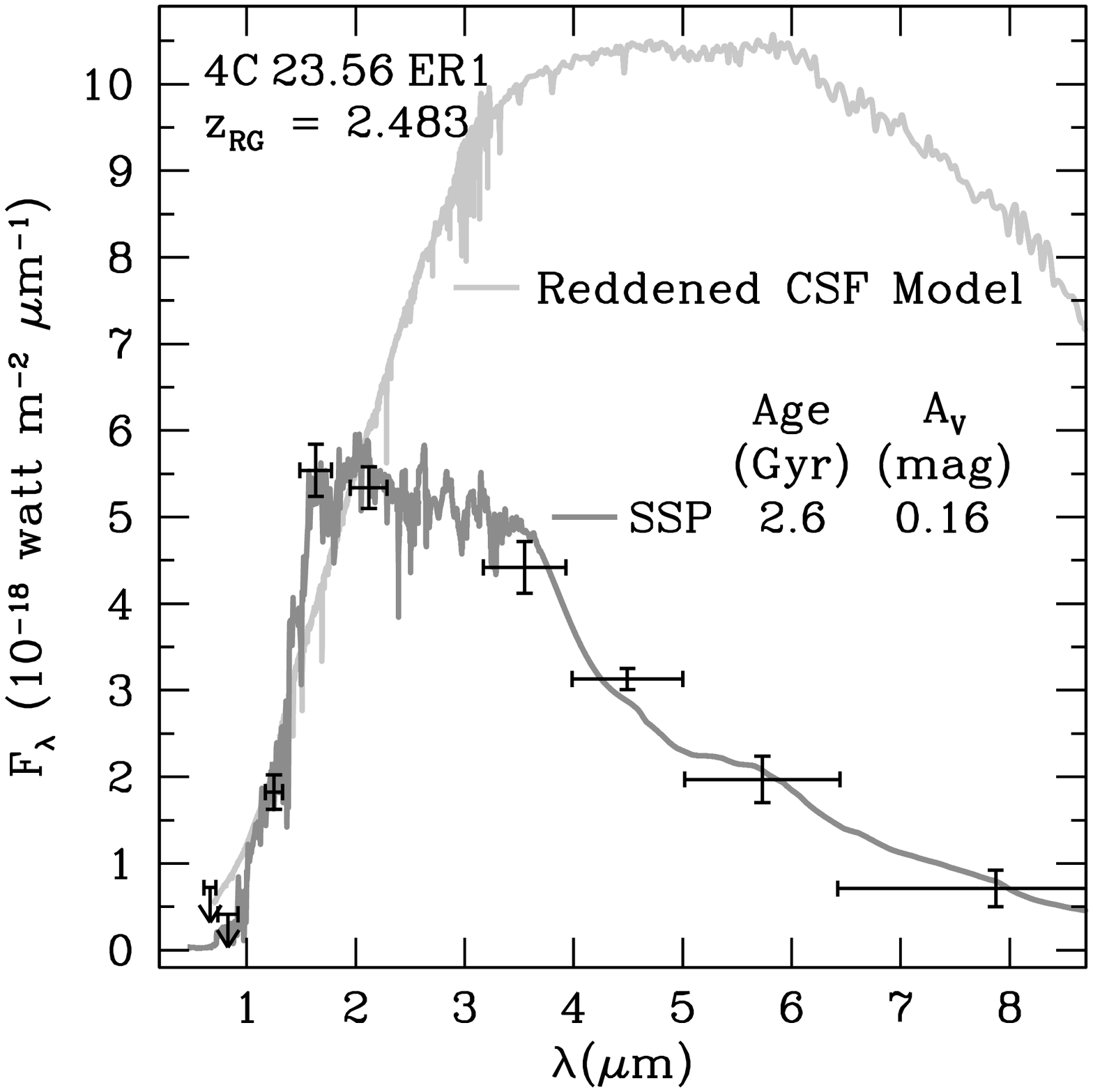}
\caption{The left panel shows the location of $z\sim2.5$ galaxies from our sample in the 
$H\!-\!K'$---$J\!-\!H$ diagram.
The galaxies are represented as points with their photometric uncertainties.
The dotted line shows the trace of solar-metallicity, instantaneous-burst \citet{bru03} models
formed at $z=10$ for redshifts ranging from 1.6 to 3.0.  The solid line traces continuous 
star-forming models, with sufficient \citet{cal00} reddening added to give $J\!-\!K'=2.9$, for the
same range of redshifts. Circled points are at $z=2.5$. The right panel shows optical, near-IR, and 
Spitzer IRAC photometry of 4C\,23.56.  The 
light gray curve is a continuous star-formation model
reddened to roughly fit the optical and near-IR data. The
dark gray curve is an extension of one of 3 SEDs given in
Stockton et al.~(2004) as fits to the optical and near-IR photometry alone. 
The IRAC points at 3.6, 4.5, 5.8, and 7.9 $\mu$m are not 
fits to the model and
have not been scaled in any way, except for the standard aperture corrections given in
the IRAC Handbook.}\label{twocol}
\end{figure}

We can explore this possibility in two ways. First, we can look at tracks of models of galaxies with
old stellar populations and of models of dusty star-forming galaxies in the $H\!-\!K'$---$J\!-\!H$
diagram. An example is shown in the left-hand panel of Fig.~\ref{twocol}. 
In this plot, the track of 
a \citet{bru03} instantaneous burst when the universe was $5\times10^8$ years old ($z\approx10$)
is plotted as a dotted line for redshifts from 1.6 to 3.0.  Similarly, continuous-star-forming models
with added dust such that $J\!-\!K'=2.9$ are shown as a solid line.  Our 6 OGs are plotted with
their photometric errors. This diagram shows two things: (1.) few, if any, of our galaxies are actually
within $2 \sigma$ of these particular dusty star-forming models; and (2.) we have probably been
missing significant numbers of galaxies that actually do have essentially unreddened old
stellar populations in our fields because of the concern about including star-forming galaxies
(i.e., all of our points are above and to the left of the OG locus).

A concern about this approach is that there can be a wide range of models for dusty 
star-forming galaxies, and we have only considered one.  It is very difficult to be sure
that {\it no} such model can match the data in this diagram.

Our second approach is simply one of extending our SED to longer wavelengths, 
such that {\it any} plausible dusty model should be distinguishable from
an old stellar population.  The first data from our Spitzer program on this sample is
shown in the right-hand panel of Fig.~\ref{twocol}.  It clearly shows that the SED of
4C\,23.56 is completely consistent with a very lightly reddened old stellar population,
and so inconsistent with our assumed dusty star-forming model that it is almost
certainly also inconsistent with any such model.

At this point, it appears that we may be stuck with the presence of massive disk galaxies of
old stars in (at least) dense environments in the early universe, along with all the 
consequences:  (1.)~the existence of massive galaxies that
have formed essentially monolithically at very early times, (2.)~rates of star formation of
at least a few hundred $M_{\odot}$ per year, sustained for several $\times10^8$ years
(or higher rates over shorter periods), and (3.)~the problem of the ultimate fate of such
galaxies, which do not seem to be common at the present epoch.

\acknowledgements 
We thank our collaborators on various phases of this program: Gabriela Canalizo, Masanori Iye, and
Toshinori Maihara.  The work has been supported in part by NSF grant AST03-073335 
and JPL contract 1278410.  It has depended on
data from the Subaru Telescope, the W. M. Keck Observatory, and the Spitzer Space
Telescope.

\end{document}